\documentclass[prb, preprint]{revtex4}
\usepackage{graphicx}
\usepackage{epsfig}
\usepackage{rotating}

\begin{document}

\title{Side-chain and backbone ordering in a Polypeptide}

\author{Yanjie Wei}
\affiliation{
Department of Physics, Michigan Technological University, 
             Houghton, MI 49931, USA}

\author{Walter Nadler}
\affiliation{Department of Physics, Michigan Technological University, 
             Houghton, MI 49931, USA}

\author{Ulrich H.E. Hansmann}
\affiliation{Department of Physics, Michigan Technological University, 
             Houghton, MI 49931, USA}
\affiliation{John-von-Neumann Institute for Computing, 
             Forschungszentrum J\"ulich, D-52425 J\"ulich, Germany}
\date{\today}

\begin{abstract}
We report results from multicanonical simulations of polyglutamic acid  chains of length ten residues. For this simple polypeptide we observe a decoupling of backbone and side-chain ordering in the folding process. While the details of the two transitions vary between the peptide in gas phase and in an implicit solvent, our results indicate that, independent of the specific surroundings, upon continuously lowering the temperature side-chain ordering occurs only after the backbone topology is completely formed.  
\end{abstract}

\maketitle

\section{Introduction}
The folding of a protein into its biologically-active structure involves a 
number of structural transitions. Examples are the collapse into an ensemble of
compact configurations, the formation of secondary structure elements such as
$\alpha$ elices or $\beta$ heets, or the ordering of side chains. The
role of these transitions in the folding process and their thermal ordering
are still only poorly understood. 
Computer experiments offer one way to study these questions but are often 
hampered by poor convergence of the simulations \cite{H05}. Only with the
development of parallel tempering \cite{PT,PT2,H97f}, multicanonical 
sampling \cite{MU,MU2}, and other generalized ensemble techniques \cite{HO98}
has it become possible to  sample efficiently in all-atom models  low-energy 
configurations of proteins with up to $\approx 50$ residues\cite{KH05}. 
This is especially important for investigations of side-chain ordering 
where one cannot fall back to the use of lattice proteins and other 
minimal protein models that usually ignore side chains.  For this reason, we have used 
multicanonical sampling \cite{MU} which was first introduced to protein
science in Ref.~\onlinecite{HO} to study  the role of side- 
chain ordering in the folding process. In order to simplify the problem
we have ignored the effects of sequence heterogeneity, i.e. side chains of 
different size. We have focused on a homopolymer, polyglutamic acid,   
as this amino acid has very long side chains that can interact through
hydrogen bonds. In that way one can expect that the effects of side chains 
on the folding process are maximized. 
Chains of a length of ten residues were simulated
both in gas phase and with an implicit solvent. 
In both cases we observed a two-step process that, 
upon lowering the temperature,
starts with a coordinated helix-formation and collapse of the polypeptide
chain. Only at much lower temperatures, after the backbone is fully organized, 
we find ordering of the side chains.  The arrangement of the side chains is due 
to the competition between the interactions among the side chains themselves 
and the interactions between them and the surrounding environment. 
Consequently,  the effect of side-chain ordering is weaker for the solvated 
molecule than for  the one in gas phase as the solvent screens the interside- 
chain interactions. However, the two-step process itself is independent  of the 
specific model indicating that the de-coupling of backbone and side-chain ordering   does not depend on the details of the  surroundings and, therefore, could be a 
common characteristic in protein folding.

\section{Methods}
Our investigations rely on simulations of $Glu_{10}$ with the ECEPP/3 force field \cite{EC} as
implemented in the 2005 version of the program package SMMP \cite{SMMP,SMMP05}.
 Here the interactions between the atoms within the homopolymer chain are approximated
 by a sum $E_{ECEPP/3}$ consisting of electrostatic energy $E_C$, a  Lennard-Jones term $E_{LJ}$, 
 hydrogen-bonding term $E_{hb}$ and a torsion energy $E_{tor}$:  
 \begin{eqnarray}
  E_{\text{ECEPP/3}} &=& E_C + E_{LJ}  + E_{hb} + E_{tor} \nonumber \\
  &=&  \sum_{(i,j)} \frac{332 q_i q_j}{\epsilon r_{ij}} \nonumber \\
 &&   + \sum_{(i,j)} \left( \frac{A_{ij}}{r_{ij}^{12}} - \frac{B_{ij}}{r_{ij}^6} \right) \nonumber \\ 
 &&   + \sum_{(i,j)} \left( \frac{C_{ij}}{r_{ij}^{12}} - \frac{D_{ij}}{r_{ij}^{10}} \right) \nonumber \\ 
  && + \sum_l U_l ( 1\pm \cos(n_l \xi_l)) \;,
\label{energy}
\end{eqnarray}
where $r_{ij}$ is the distance between the atoms $i$ and $j$,  $\xi_l$ is the $l$th torsion 
angle, and energies are measured in kcal/mol.  
The protein-solvent interactions are approximated by a solvent accessible surface term 
\begin{equation}
  E_{solv} = \sum_i \sigma_i A_i \;.
\label{solventEnergy}
\end{equation}
The sum goes over the  solvent accessible areas $A_i$ of all atoms $i$ weighted by solvation
parameters $\sigma_i$ as determined in Ref.~\onlinecite{OONS}, a common choice when the 
ECEPP/3 force field is utilized. Note that $E_{solv}$ is a rather crude approximation of the
interaction between the polypeptide and the surrounding water that is motivated by the
low computational costs when compared to simulations with explicit water molecules. 
Because of that its reliability is not
always clear \cite{PTH}. Especially the temperature scale may be distorted (leading, for instance,
to transitions at temperatures where in nature water would be vaporized). However, our
previous experiences \cite{PTH,TTH} have shown that our energy function reproduces 
{\it qualitatively} the  effects of protein-water interaction correctly.

The evaluation of this detailed energy function is not only computationally more expensive
than that of minimal protein models but the competing interactions lead also to an
energy landscape that is characterized by a multitude of minima separated by high
energy barriers. As the probability to cross an energy barrier of height $\Delta E$ is given by
$\exp (-\Delta E/k_BT)$ ($k_B$ the Boltzmann constant) it follows that  extremely long runs are necessary to obtain sufficient statistics in regular canonical simulations at a low temperature $T$.
Hence, in order to enhance sampling,  we rely on the multicanonical approach \cite{MU,MU2} as
described in Ref.~\onlinecite{HO}.  Here, configurations are weighted with an iteratively
determined term $w_{MU} (E)$ such that the probability distribution
\begin{equation}
    P_{MU}(E) \propto n(E) w_{MU}(E) \approx const~,
\end{equation}
where $n(E)$ is the spectral density of the system. Thermodynamic averages $<O>$ at temperature $T$
are obtained by re-weighting \cite{FS}: 
\begin{equation}
 <O>(T) = \frac{\int dx \; O(x) e^{-E(x)/k_BT} / w_{MU}[E(x)]}
                          {\int dx \; e^{-E(x)/k_BT} / w_{MU}[E(x)]}
\end{equation}
where $x$ counts the configurations of the system.

After  determining the multicanonical weights $w_{MU}(E)$,  we have performed  multicanonical
simulations of $5\times 10^6$  sweeps. Each sweep consists of 70 Metropolis steps that try to update
each of the 70 dihedral angles (the degrees of freedom in our system) once. Every 10 steps
various quantities are measured and written to a file for further analysis. These include 
the energy $E$ with its respective contributions from Eq.~(\ref{energy})
and - in the case of the simulations in solvent - from the protein-solvent interaction energy $E_{solv}$.
The radius of gyration $r_{gy}$ is a measure of the 
geometrical size, and the number of helical residues $n_H$, i.e. residues where the pair of dihedral angles
$(\phi,\psi)$ take  values in the range 
($ -70^\circ \pm 30^\circ $, $ -37^\circ \pm 30^\circ $) \cite{Okamoto1995}.
Finally, we monitor the number of hydrogen bonds, $n_{HB}$ where we distinguish between 
hydrogen bonds along the backbone and hydrogen bonds between side chains. 

\section{Results and Discussions}
Our aim is to study the relationship between side-chain ordering and other transitions for
the example of the homopolymer Glu$_{10}$. We first investigate the case of molecules in
gas phase. Fig.~\ref{sph_gp} displays the specific heat per molecule 
\begin{equation}
     C(T) = k_B \beta^2 (<E^2> - <E>^2)
\label{specHeat}
\end{equation}
as a function of temperature. Two peaks are observed in this plot indicating two transitions.

The first peak is located at a temperature  $T_1=590$ K and is relatively broad, with a half-width of about 160K.
The corresponding plot of the helicity in Fig.~\ref{nH_gp} shows that this peak separates
a high-temperature region where the backbone has no ordering from a region where temperatures
are low enough to allow the formation of backbone hydrogen bonds (see the inlay to Fig.~\ref{nH_gp})
and subsequent growth of an $\alpha$-helix.  While the width of the specific heat peak is comparable to that of Ala$_{10}$  \cite{Hansmann1999}, the transition temperature here is considerably higher ($T_1=427$ K for Ala$_{10}$).   Since the side chains of Glu are larger than in the case of Ala, they 
provide sterical hindrances to backbone conformations, leading to a decrease of the backbone entropy.
As the transition is driven  by entropy, this  leads to a higher transition temperature.
In general, it is not uncommon that transition temperatures in the gas phase are relatively high
\cite{Hansmann1999} and it has been verified experimentally that helices can be stable in the gas phase up to high temperatures \cite{Jarrold2003b,Jarrold2004}.

The helix-coil transition is also connected with a collapse of the molecule. This can be seen
from the inlay of Fig.~\ref{sph_gp}  where we display the average radius of gyration $r_{gy}$ as a function of temperature. We  observe
a monotonic drop in the radius of gyration, starting already at higher temperatures but continuing
through the transition regime as defined by the specific heat peak. Below the transition
$r_{gy}$ stabilizes, reflecting the stable helical structure that has been reached.

The inlay 
also shows that the second peak in the specific temperature that is observed
at a lower temperature $T_2=164$ K in Fig.~\ref{sph_gp} is not related to the collapse of the molecule. 
Hence, it is reasonable to assume that this transition is related to an ordering of the side chains in our molecule. 
This hypothesis is supported by  Fig.~\ref{sc_gp}
where we display the average number $n_{hb}^S$ of side-chain hydrogen bonds as function of temperature.  The fluctuations of this quantity, $\chi(T)=\left<\left(n_{hb}^S-\left<n_{hb}^S\right>\right)^2\right>$, are shown in the inlay.
The change in the number of side-chain hydrogen-bonds that is observed in this figure at $T_2$ clearly shows that the corresponding peak in the specific heat indicates indeed a second transition that separates now a low-temperature phase with ordered side chains from a phase at temperatures above $T_2$ where the backbone is ordered but the side chains are still fluctuating. The form of the side-chain ordering can be seen best from the lowest energy conformation displayed in Fig.~\ref{gmc_gp}.  
Here, as also already described earlier in Ref.~\onlinecite{Scheraga68},  the side chains nestle 
along the cylinder formed by the helix  and are stabilized by the side chain hydrogen bonds 
(not shown in the figure).  
Hence, our results indicate that Glu$_{10}$ "folds" in gas phase in a two-step process that - upon continuously lowering the temperature - starts
with a concurrently occurring collapse and secondary structure formation. 
Only after  the backbone geometry is fixed can the side chains align themselves, too, in a second step.

In nature, proteins are solvated and the details of the protein-solvent interaction are important for
the structure and function of a protein. Hence, it is not clear whether our results obtained in
gas phase apply also for solvated proteins. For this reason we have extended our investigation 
in a second step to that of solvated Glu$_{10}$.  As in the case of the molecule in gas phase,
we observe again two peaks in the specific heat (see Fig.~\ref{sph_s}). The peak at the
higher temperature $T_1=477$ K   marks again the collapse (see the inlay of Fig.~\ref{sph_s}) and 
subsequent formation of an $\alpha$-helix. The later can be seen from Fig.~\ref{nH_s} where we display again
the average helicity and average number of backbone hydrogen bonds.
Note that the transition temperature is by more than 100 K lower than in the gas phase. Also. the
peak in the specific heat is higher and narrower, indicating a sharper, more well-defined
transition. Both features are actually known from earlier studies on polyalanine 
\cite{Hansmann1999,Peng2003}. The reason for the shift to a lower temperature is the competition between the formation of backbone 
hydrogen bonds that stabilize an $\alpha$ helix, and that of hydrogen bonds between the backbone and the solvent in the coil phase,
the energetic contribution of the latter being 
described in a mean field way by the solvent energy term (\ref{solventEnergy}). While the
transition in gas phase is driven solely by entropy, here also a part of the energy
favors the coil phase. These effects collaborate so that the transition takes
place at a lower temperature and becomes sharper. 
The sharper transition is also particularly visible in the radius of gyration,
shown in the inlay of Fig.~\ref{sph_s}.
In contrast to the gas phase, both quantities remain practically constant above the transition. At $T_1$, however,
both show a sharp drop and remain practically constant again in the ordered phase.

The actual value of the transition temperature is still
higher than that of the corresponding value for $Ala_{10}$ ($T_1=333$ K). 
The absolute differences between the transition temperatures of both molecules, $Ala_{10}$ and $Glu_{10}$,  are relatively similar, 163 K  in vacuum and 144 K in solvent. The reasons discussed for this shift in vacuum appear to apply here in the solvent case, too:
Lower backbone entropy due to larger side chains.
We remark that  experimental measurements would allow for an 
adjustment of the temperature scale which seems to be incorrect for the ECEPP/3 force field.

The second peak in the specific heat that is observed in Fig.~\ref{sph_s} at the lower temperature $T_2 = 111$ K  is more narrow and smaller than the corresponding one for the molecule in gas phase (Fig.~\ref{sph_gp}).   As with the coil-helix transition temperature $T_1$, 
this transition is also shifted to lower temperatures in the solvent, albeit by the smaller amount of 53 K. As in the case of  of Glu$_{10}$ in gas phase, the side chains are ordered at temperatures below $T_2$.  This can be seen for the example of the lowest energy configuration that was found in our simulation of the solvated Glu$_{10}$ molecule. This structure is shown in Fig.~\ref{gmc_s} and is characterized by side chains that order themselves by extending into the solvent. This is in strong contrast to the gas phase, where the the side chains nestle along the helical cylinder and the ordering results from hydrogen bond  
formation between the side chains. Practically no side-chain hydrogen bonds  are observed in the low temperature phase ($\left< n_{hb}^S\right> < 0.04 $) for the solvated molecule. 

No correlation is observed between the number of side-chain hydrogen bonds
and the peak of the specific heat at $T_2$ in
Fig.~\ref{sph_s}.  Hence, the mechanism  by which the side chains order themselves has to be 
different in water from  in gas phase. The radial orientation of the side chains in the lowest energy configuration of Fig.~\ref{gmc_s} suggest that they try to enhance exposure to water. As Glu$_{10}$ is a hydrophilic molecule such behavior would be reasonable as it would decrease the solvation energy. Although no pronounced decrease in the solvation energy can be seen around
$T_2$,  the $fluctuations$ of the solvation energy show a pronounced increase, see Fig.~\ref{solvefluct}.
Note that in this figure the fluctuations are normalized by temperature, $<\delta E^2>/k_BT^2$,
to allow a comparison with the specific heat (\ref{specHeat}).
These solvent energy fluctuations at $T_2$ are actually even larger than at the helix-coil transition.
There the energy fluctuations and, hence, the specific heat 
are dominated by the fluctuations of the internal energy. In addition, at the helix-coil transition
the fluctuations of both energy contributions exhibited a peak at the transition, together
with a $negative$ peak of the cross correlation, the latter denoting that
the fluctuations  of internal and solvent energies are anti-correlated.
In contrast, around $T_2$ we observe an increase of the solvent energy fluctuations that continues to lower temperatures, while the fluctuations of the internal energy show a plateau and decrease below $T_2$. The peak in the specific heat around $T_2$ is actually due to increasing anticorrelations below $T_2$. The latter reflects the fact that in the side-chain ordered phase fluctuations in the solvent energy are countered by corresponding negative fluctuations in the internal energy.

Hence, while the structure of the solvated molecule also evolves in a two-step process upon continuously lowering the temperature, the mechanism that leads to the second step, the side-chain ordering, is different from the gas phase. There the ordering of side chains was due to formation of hydrogen bonds between the side chains. 
However, in water, the polar side chains interact directly with the surrounding water. Water screens them from forming hydrogen bonds among each other.

\section{Summary and Outlook}

Our results indicate that Glu$_{10}$ "folding" in gas phase 
is a two-step process, starting - upon continuously lowering the temperature  - with a collapse which becomes concurrent with the secondary structure formation.
Only after the backbone geometry is fixed the side chains can align themselves along the helical cylinder, too, stabilizing themselves by
forming hydrogen bonds with each other.

In solvent, Glu$_{10}$ "folds" also in a two-step process upon continuously lowering the temperature. However, in contrast to the above scenario, the collapse is concurrent with the secondary structure formation that exhibits a much sharper transition than in vacuum. Side-chain ordering takes place, too, but it has a different character. Side chains do not align themselves along the helical cylinder but rather, extend into the solvent, which screens them from forming hydrogen bonds among themselves. 

 Our results indicate that the de-coupling of backbone and side-chain ordering  does not 
depend on the details of the environment. Hence, it is reasonable to assume that this process
could be a common characteristic in protein folding. In order to test this hypothesis, we are
now looking at the dependence of side-chain ordering on size and chemical properties. Glutamine (Gln), for example, 
is about the size of glutamic acid and should also be able to participate in hydrogen bonds.
Aspartic acid (Asp) and asparagine (Asn) are also able to form hydrogen bonds
but have a smaller size,
while lysine (Lys) has a larger polar side chain.
Investigations of these molecules along the lines sketched in this contribution will add more detailed knowledge
to side chain ordering in polypeptides. \\ \\

{\em Acknowledgments }
We thank H. Scheraga for discussions and for drawing our attention to 
Ref.~\onlinecite{Scheraga68}.  Support  by a
research grant (CHE-0313618) of the National Science Foundation (USA) is acknowledged.


%
%
%
\clearpage
{\huge Figure captions:}

\begin{description}

\item{Fig.~1}  Specific heat $C(T)$ as function of temperature $T$  for  Glu$_{10}$ in gas phase as obtained from a multicanonical simulation with $5\times 10^6$ sweeps. The inlay shows the
average radius of gyration  $<r_{gy}>(T)$ .

\item{Fig.~2} Average number of helical residues $<n_H>(T)$ as function of temperature $T$ for Glu$_{10}$ in gas phase as obtained from a multicanonical simulation with $5\times 10^6$  sweeps. The inlay shows the corresponding average number $<n_{hb}^B>(T)$ of backbone hydrogen bonds as function of temperature $T$.

\item{Fig.~3:} Average number of side chain hydrogen bonds $<n_{hb}^S>(T)$ as function of temperature $T$ for  Glu$_{10}$ in gas phase as obtained from a  multicanonical simulation  with $5\times 10^6$  sweeps. The inlay shows the corresponding fluctuation $\chi (T)$  as function of temperature $T$.

\item{Fig.~4:} Lowest energy configuration of  Glu$_{10}$ in gas phase as obtained from a  multicanonical simulation  with $5\times 10^6$ sweeps  and subsequent minimization. The picture has been obtained with PYMOL.

\item{Fig.~5:} Specific heat $C(T)$ as function of temperature $T$ for solvated  Glu$_{10}$ as obtained from a multicanonical simulation with $5\times 10^6$  sweeps. The inlay shows the average radius of gyration  $<r_{gy}>(T)$ .

\item{Fig.~6} Average number of helical residues $<n_H>(T)$ as function of temperature $T$ for  the solvated Glu$_{10}$ as obtained from a multicanonical simulation  with $5\times 10^6$ sweeps. The inlay shows the corresponding average number $<n_{hb}^B>(T)$ of backbone hydrogen bonds as function of temperature $T$.

\item{Fig.~7:} Lowest energy configuration of the solvated  Glu$_{10}$ as obtained from a  multicanonical simulation  with $5\times 10^6$ sweeps and subsequent minimization. The picture has been obtained with PYMOL.

\item{Fig.~8:}  Fluctuations of the solvation energy $<\delta E^2_{solv}>(T)$  and intramolecular energy $<\delta E^2_{ECEPP/3}>(T)$ for solvated  Glu$_{10}$  
 as obtained from a multicanonical simulation with $5\times 10^6$  sweeps. In addition,
 the cross-correlation $<\delta E_{solv} \delta E_{ECEPP/3}>(T)$ is shown.
 Note that $\delta E = E - <E>$.
\end{description}
\setcounter{figure}{0}
\clearpage
\begin{sidewaysfigure}
    \includegraphics[width=1.0\columnwidth]{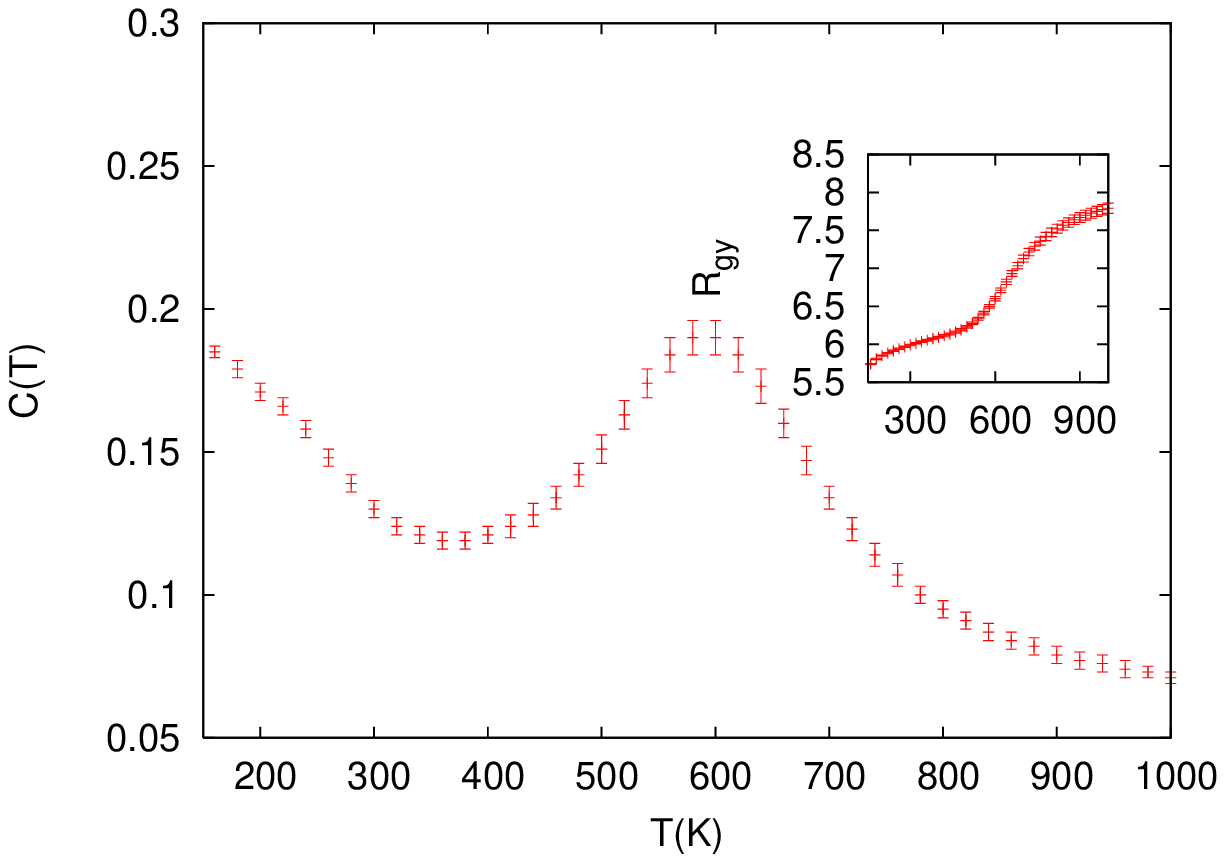}
\caption{\label{sph_gp}}
\end{sidewaysfigure}
%
\clearpage
\begin{sidewaysfigure}
    \includegraphics[width=1.0\columnwidth]{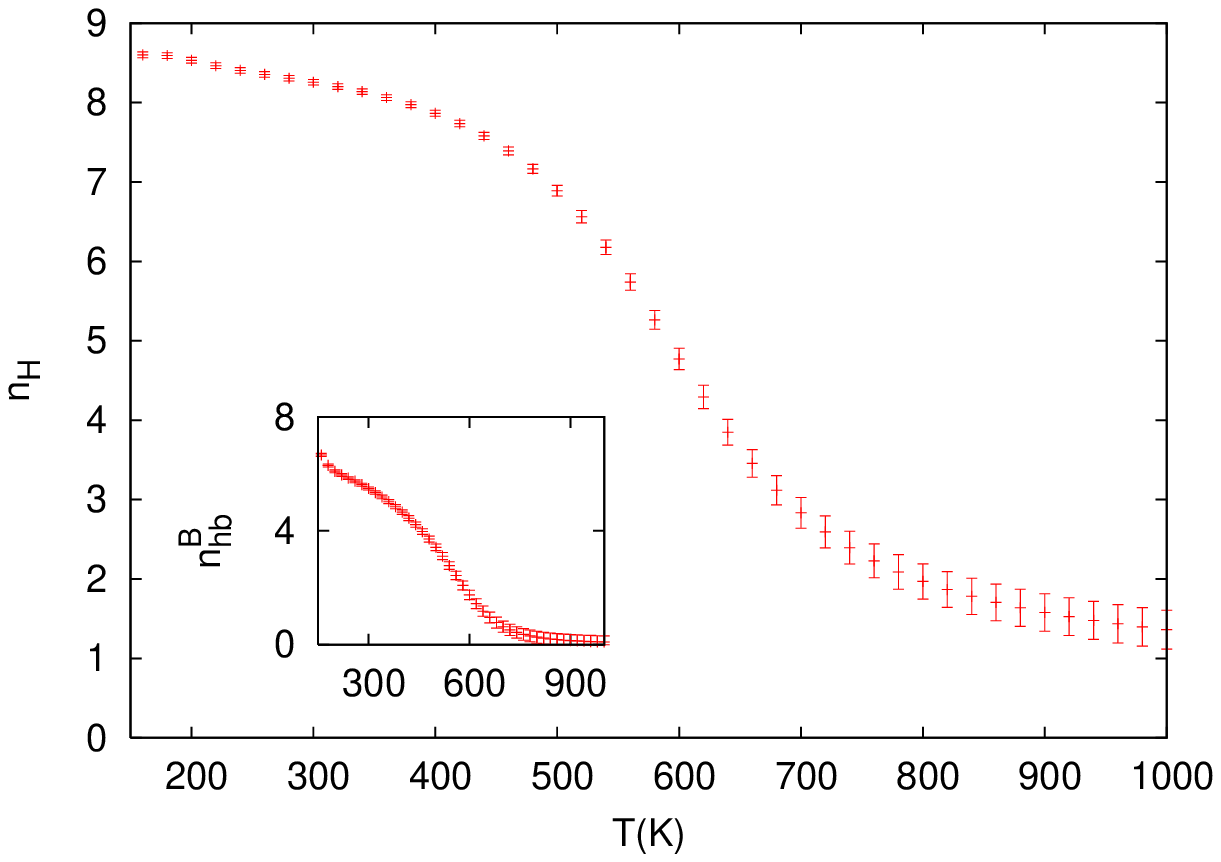}
\caption{\label{nH_gp}}
\end{sidewaysfigure}
%
\clearpage
\begin{sidewaysfigure}
  \includegraphics[width=1.0\columnwidth]{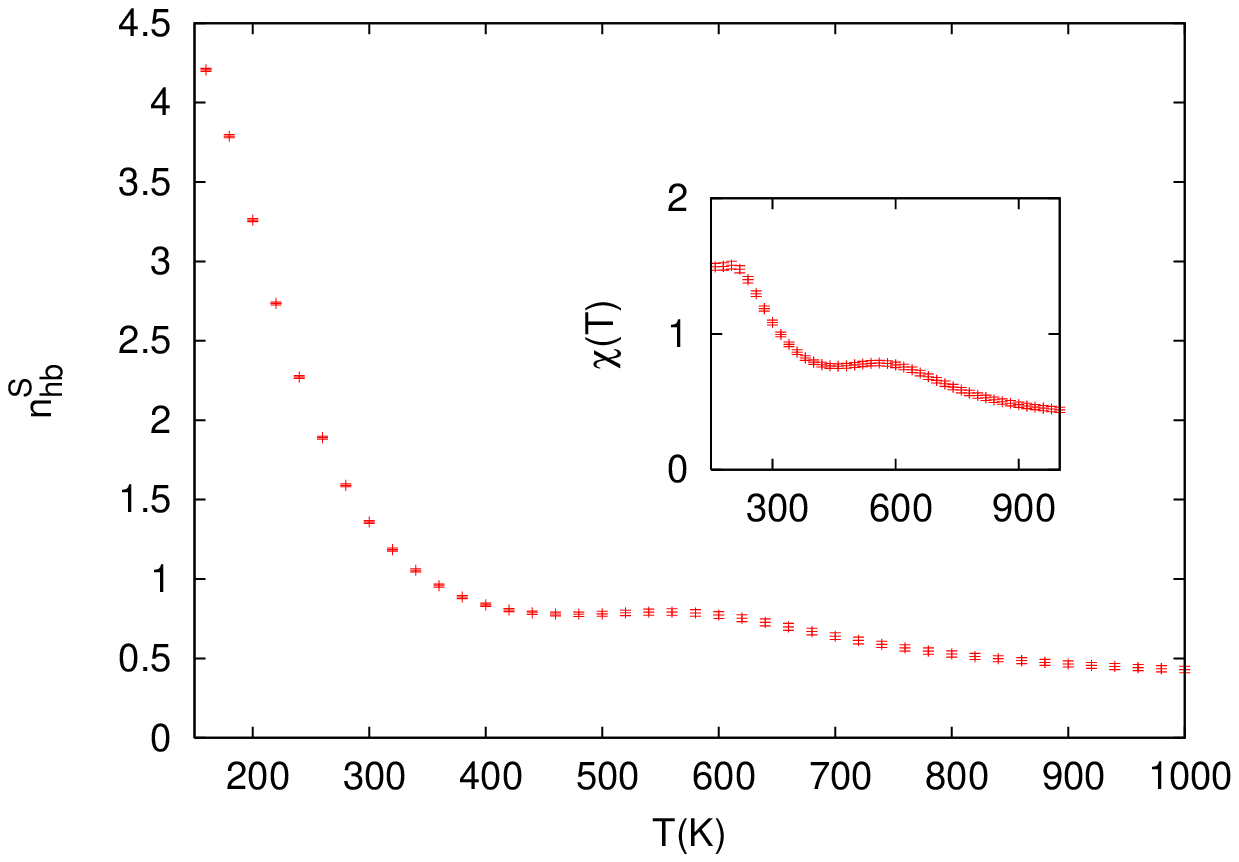}
\caption{\label{sc_gp}}
\end{sidewaysfigure}
%
\clearpage
\begin{sidewaysfigure}
   \includegraphics[width=1.0\columnwidth]{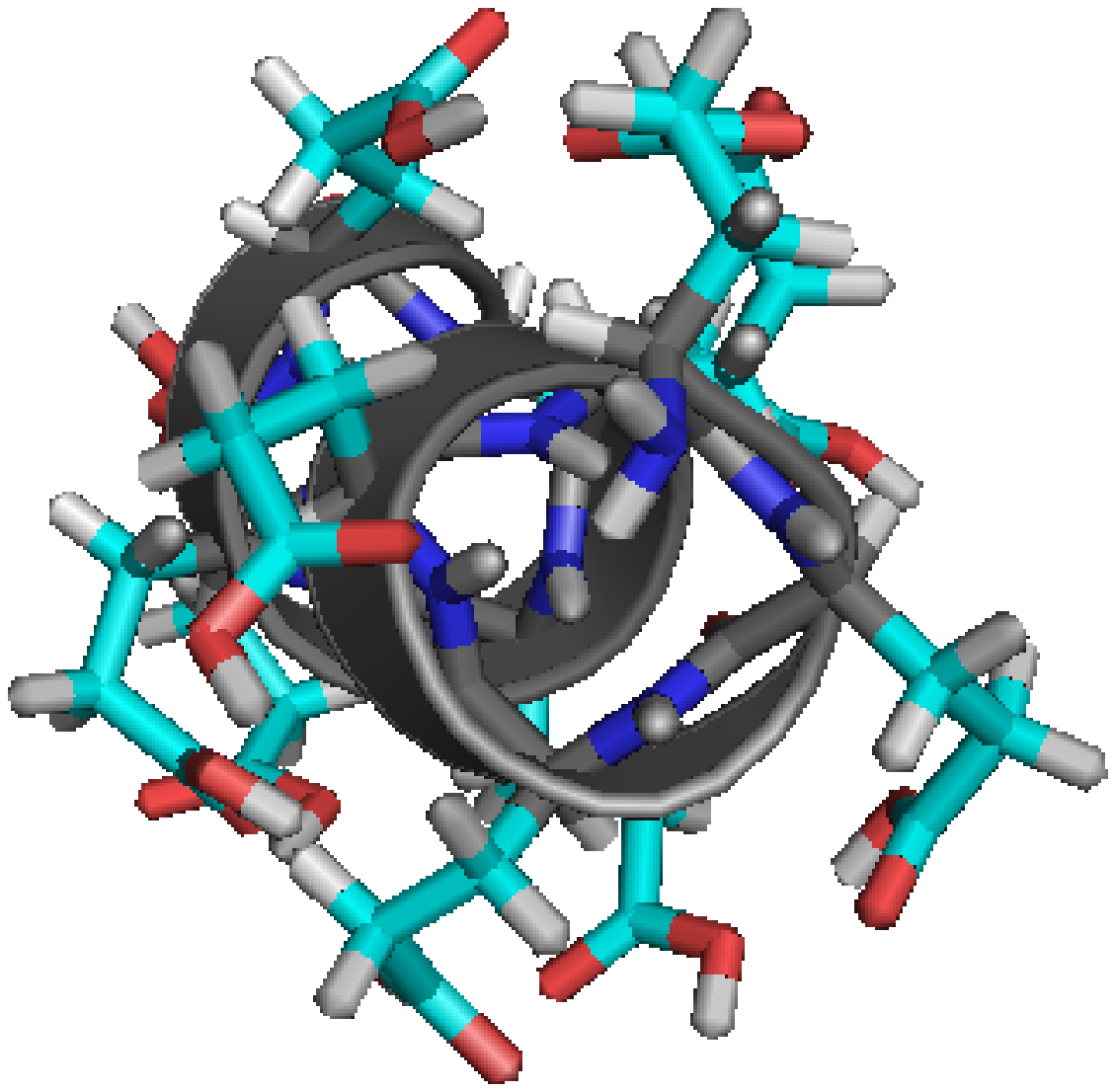}
\caption{\label{gmc_gp}}
\end{sidewaysfigure}
%
%
\clearpage
\begin{sidewaysfigure}
   \includegraphics[width=1.0\columnwidth]{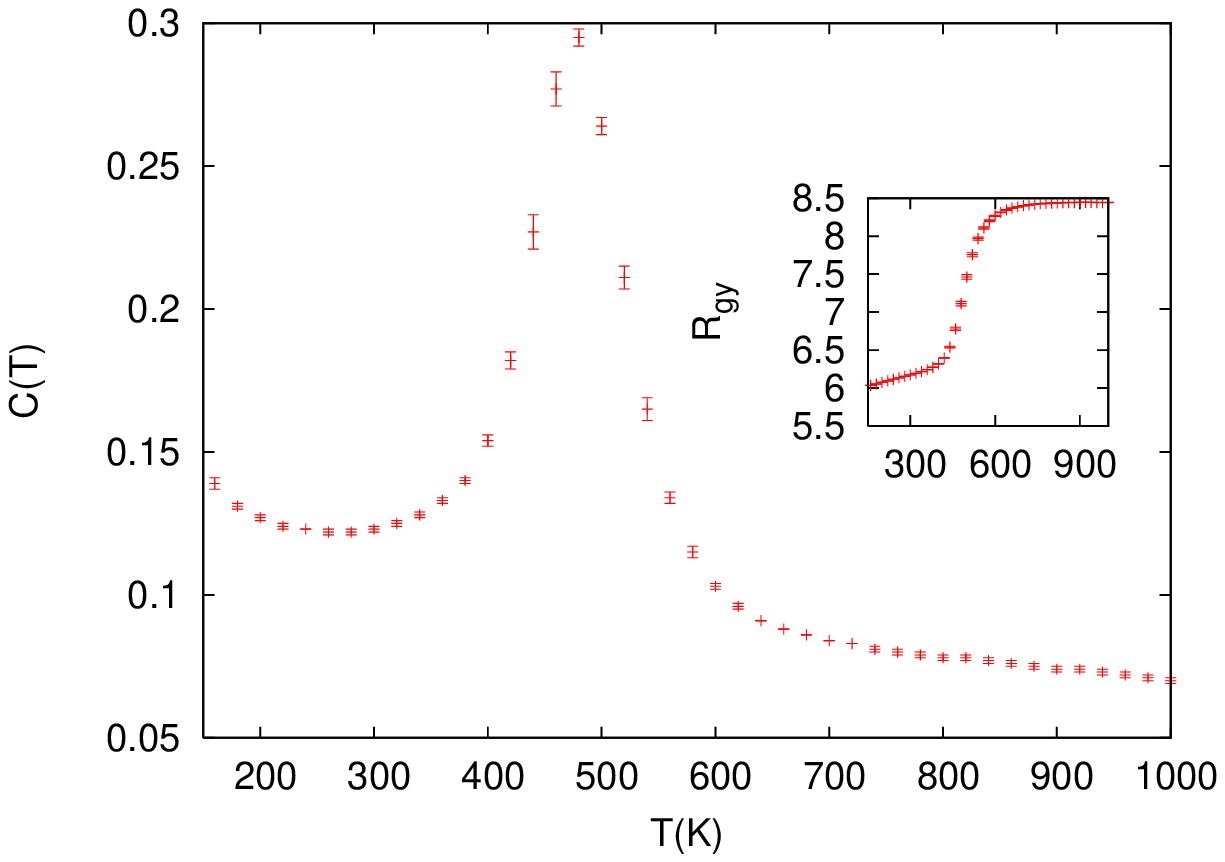}
\caption{\label{sph_s}}
\end{sidewaysfigure}
%
\clearpage
\begin{sidewaysfigure}
   \includegraphics[width=1.0\columnwidth]{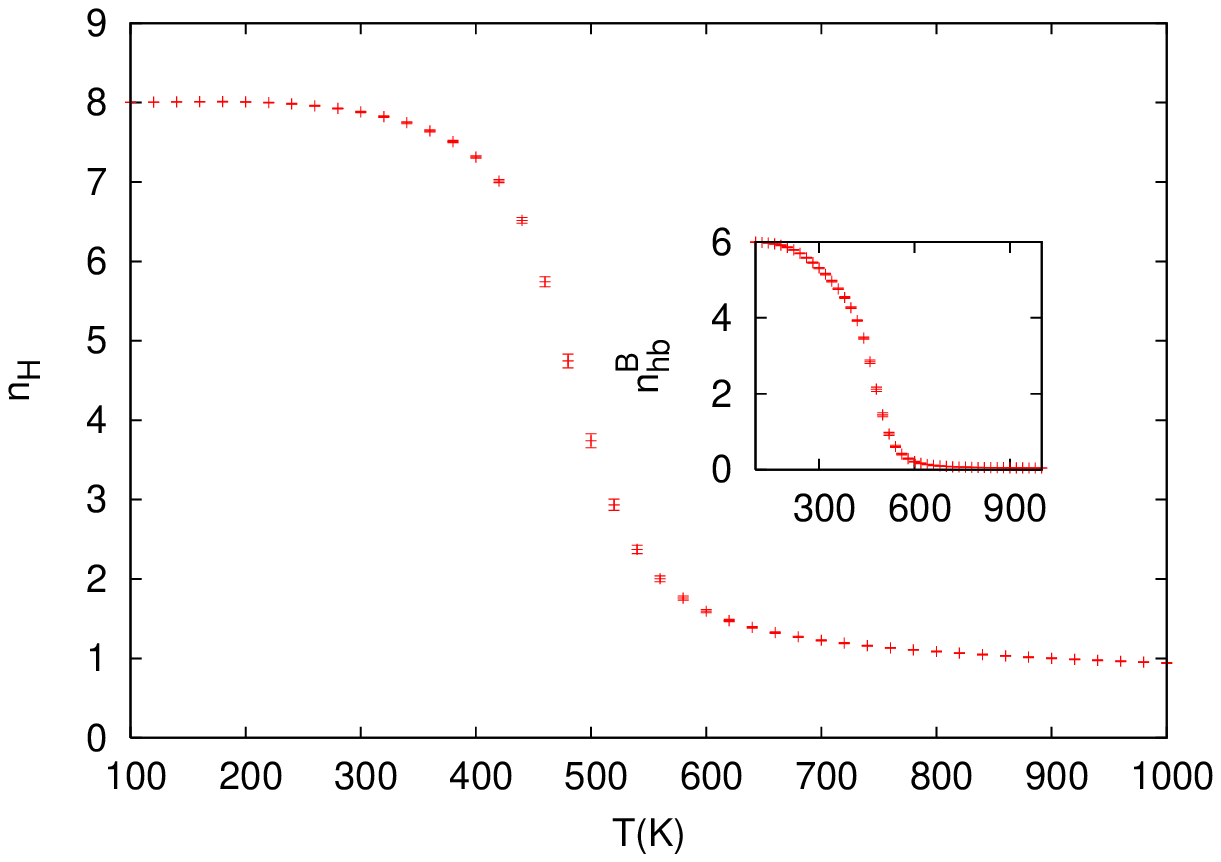}
\caption{\label{nH_s}}
\end{sidewaysfigure}
\clearpage
\begin{sidewaysfigure}
   \includegraphics[width=1.0\columnwidth]{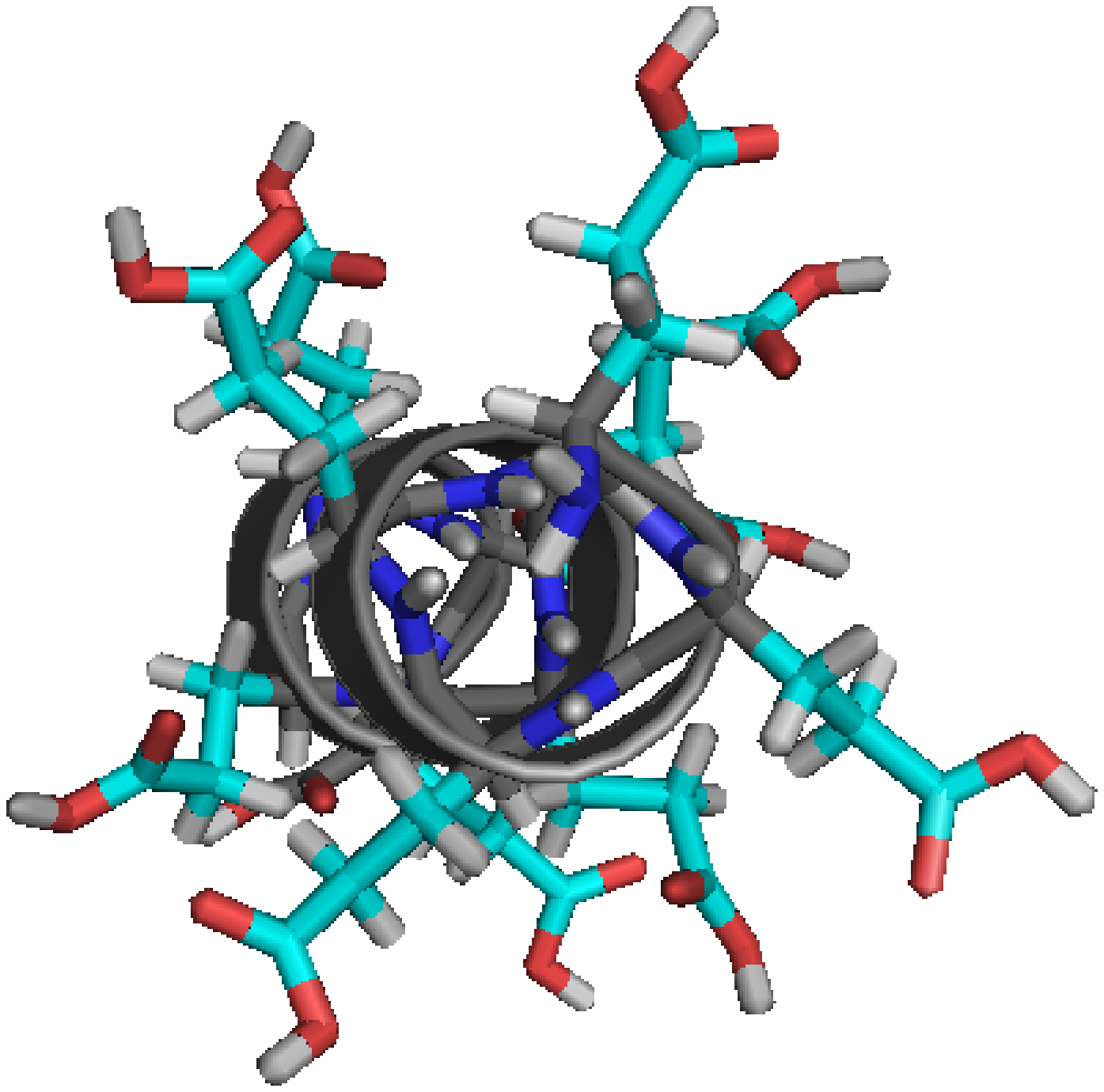}
\caption{\label{gmc_s}}
\end{sidewaysfigure}
%
\clearpage
\begin{sidewaysfigure}
   \includegraphics[width=1.0\columnwidth]{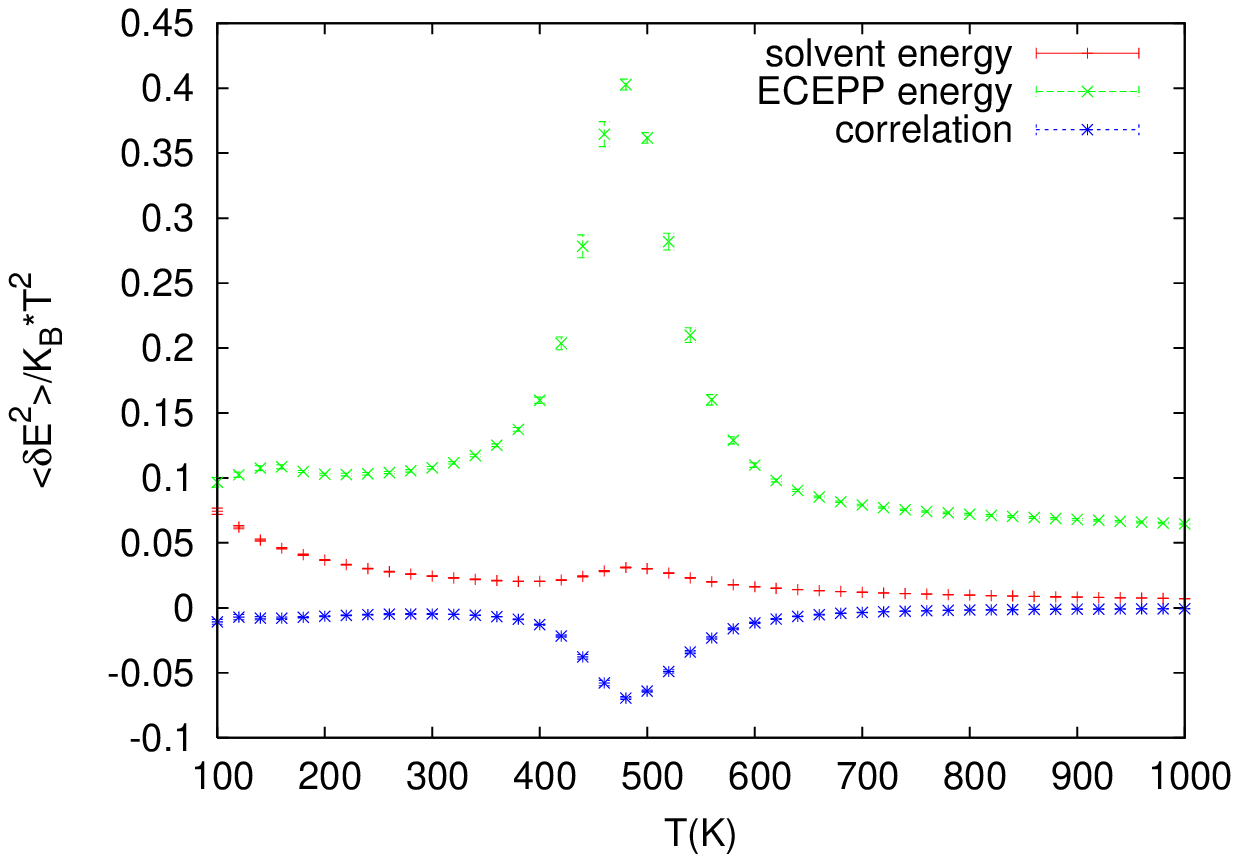}
\caption{\label{solvefluct}}
\end{sidewaysfigure}
\end{document}